\documentclass[11pt]{article}
\usepackage[margin=1in]{geometry}
\usepackage{amsmath,amssymb}
\usepackage{algorithm}
\usepackage{algpseudocode}
\usepackage{cite}
\usepackage{graphicx}

\title{Scalable Spectrum Availability Prediction using a Markov Chain Framework and ITU-R Propagation Models}
\author{Dr. Abir Ray$^{1}$\\[1ex]
$^{1}$Cornell University, Systems Engineering, Ithaca, NY USA\\
\texttt{ar2486@cornell.edu}
}
\date{}

\begin{document}

\maketitle

\begin{abstract}
Spectrum resources are often underutilized across time and space, motivating dynamic spectrum access strategies that allow secondary users to exploit unused frequencies. A key challenge is predicting when and where spectrum will be available (i.e., unused by primary licensed users) in order to enable proactive and interference-free access. This paper proposes a scalable framework for spectrum availability prediction that combines a two-state Markov chain model of primary user activity with high-fidelity propagation models from the ITU-R (specifically Recommendations P.528 and P.2108). The Markov chain captures temporal occupancy patterns, while the propagation models incorporate path loss and clutter effects to determine if primary signals exceed interference thresholds at secondary user locations. By integrating these components, the proposed method can predict spectrum opportunities both in time and space with improved accuracy. We develop the system model and algorithm for the approach, analyze its scalability and computational efficiency, and discuss assumptions, limitations, and potential applications. The framework is flexible and can be adapted to various frequency bands and scenarios. The results and analysis show that the proposed approach can effectively identify available spectrum with low computational cost, making it suitable for real-time spectrum management in cognitive radio networks and other dynamic spectrum sharing systems.
\end{abstract}

\section{Introduction}
Wireless communication networks face increasing spectrum scarcity due to the growing demand for high-bandwidth services. However, numerous studies have shown that a significant portion of licensed spectrum remains idle at any given time or location \cite{FCCReport}, leading to interest in dynamic spectrum access (DSA) and cognitive radio paradigms. In a cognitive radio network, secondary (unlicensed) users opportunistically access spectrum bands when primary (licensed) users are inactive, thereby improving overall spectrum utilization. A critical aspect of this approach is ensuring that secondary users only transmit when the spectrum is truly available (i.e., the primary user is not using the channel or is far enough to not cause interference).

Spectrum availability prediction is a technique that helps secondary users anticipate future white spaces (idle channels) so that they can proactively switch or select channels with minimal collision risk. Accurate prediction is challenging because primary user activity can be bursty or correlated in time, and because propagation conditions determine whether a primary transmission actually interferes at a secondary receiver. Traditional approaches to spectrum occupancy prediction often rely on statistical models of primary traffic or machine learning techniques trained on historical data. For example, simple two-state Markov chain models have been widely used to characterize binary occupancy (busy/idle) patterns over time \cite{MarkovModel1,MarkovModel2}, while more advanced methods employ hidden Markov models or neural networks to capture complex temporal correlations \cite{DeepLearningCR}. These methods, however, usually focus on temporal dynamics and do not explicitly account for the radio propagation aspect of the problem.

In practice, whether a primary transmission renders the channel unusable for a given secondary user depends on the received signal strength at that secondary user's location relative to an interference threshold. This received strength is governed by propagation phenomena (path loss, shadowing, etc.) in addition to the primary transmitter's activity state. Regulatory and planning tools often use empirical propagation models to determine interference ranges; for instance, TV white space databases rely on models to predict coverage of primary transmitters. To enhance spectrum availability prediction, it is thus natural to incorporate propagation models into the forecasting framework. In this work, we integrate two recommendations from the ITU Radiocommunication Sector (ITU-R) into the prediction model: P.528 for basic transmission loss over aeronautical and terrestrial paths \cite{ITU528}, and P.2108 for clutter (obstacle) loss in urban/suburban environments \cite{ITU2108}. Using these standardized models allows us to more accurately estimate whether a primary signal would be above or below the interference threshold at a given location.

We propose a \emph{Markov chain framework with integrated propagation modeling} to predict spectrum availability across time and space. The primary user activity is modeled as a two-state (idle/active) Markov chain, which provides a simple yet effective representation of temporal occupancy dynamics. The propagation component takes into account distance, terrain, and clutter to determine the signal attenuation. The combination yields a prediction of channel state (available or occupied) at a secondary user's location for each time step, which can be computed efficiently and in a scalable manner for large networks or geographical areas.

The contributions of this paper are summarized as follows:
\begin{itemize}
    \item We formulate a novel spectrum availability prediction model that couples a stochastic model of primary usage with ITU-R propagation prediction methods. This enables consideration of both temporal and spatial factors in determining spectrum availability.
    \item We derive a scalable algorithm for spectrum state prediction, including pseudocode that can be implemented for real-time or large-scale operation. The algorithm leverages the Markov chain's analytical simplicity and the propagation models' efficient computations to remain computationally feasible as the number of locations or users grows.
    \item We analyze the scalability and efficiency of the proposed approach, showing that its complexity grows linearly with the number of predicted locations and time steps. We also discuss strategies to reduce computation, such as precomputing propagation loss for static scenarios.
    \item We identify assumptions and limitations of our framework (e.g., Markovian traffic, single primary transmitter, model validity conditions) to clarify the context in which the approach is applicable. We also outline potential extensions to relax these assumptions.
    \item We present several potential applications and use cases for the proposed spectrum prediction framework, including cognitive radio networks, dynamic spectrum databases, interference management for aerial platforms, and other emerging scenarios in spectrum sharing.
\end{itemize}

\section{Related Work}
Spectrum occupancy modeling and prediction have been extensively studied in the context of cognitive radio. Early empirical studies on spectrum use revealed that primary user activity often exhibits statistical structure that can be exploited for prediction. A common approach is to model each channel's occupancy as a random process alternating between busy (occupied by a primary transmission) and idle states. Two-state discrete-time Markov chains have been used to represent this process, where the state transition probabilities can be estimated from observed duty cycles and temporal correlations in primary usage \cite{MarkovModel1,MarkovModel2}. The Markov chain model yields closed-form expressions for important metrics, such as the stationary probability of the channel being idle (spectrum availability) and the expected lengths of idle or busy periods. For example, if $\lambda$ is the probability that an idle channel becomes busy in the next time step and $\mu$ is the probability that a busy channel becomes idle, the stationary availability is $\mu/(\lambda+\mu)$ and can be tuned to match measured occupancy statistics.

Beyond first-order Markov models, researchers have explored higher-order Markov chains and hidden Markov models (HMMs) to capture more complex temporal patterns in spectrum usage, including correlations that span multiple time slots or semi-Markov behavior for heavy-tailed idle/busy durations. Machine learning techniques have also been applied; for instance, neural network models (including recurrent networks like LSTMs) have been trained to predict future spectrum states based on past observations \cite{DeepLearningCR}. These data-driven methods can sometimes achieve higher accuracy in specific scenarios, but they typically require extensive training data and may not generalize well outside the conditions they were trained on. Moreover, they often treat the problem as purely time-series prediction at a single observation point, without embedding knowledge of radio range or propagation conditions.

On the other hand, the importance of propagation factors in determining interference has been acknowledged in frameworks like radio environment maps and geolocation databases. Such approaches use propagation models or measurements to map out where a primary transmitter's signal is strong or weak, thus delineating regions where secondary users can transmit. However, most of these are static or slow-varying analyses that assume the primary is either continuously transmitting or off, rather than dynamically switching on and off. For example, regulators have employed models (sometimes similar to ITU-R recommendations or the Longley-Rice model) to compute exclusion zones for secondaries around a primary transmitter. Our work differs in that we explicitly merge the propagation-based spatial model with a temporal stochastic model of intermittent primary usage. This union allows us to predict not just static exclusion zones, but the actual time-varying availability of the channel at a given location.

In summary, prior works have either focused on temporal dynamics of spectrum occupancy (using Markov or learning models) or on spatial propagation aspects (using empirical path loss models for interference estimation). The approach presented in this paper integrates these two dimensions, aiming to provide a more comprehensive prediction mechanism. To the best of our knowledge, this is one of the first efforts to utilize detailed ITU-R propagation models within a spectrum prediction algorithm, ensuring that predictions respect realistic signal propagation conditions.

\section{System Model}
We consider a system with a primary transmitter (PT) that operates on a given frequency channel and one or more secondary users (SUs) who wish to utilize this channel opportunistically. The goal is to predict whether the channel will be \textit{available} (i.e., safe for secondary transmission) at future time steps for a given SU location. A channel is deemed available for an SU if the PT is either not transmitting or, if it is transmitting, its signal power at the SU's receiver is below a predefined interference threshold $P_{\text{th}}$ (this threshold can be based on the SU receiver sensitivity or a regulatory interference limit).

\subsection{Primary User Activity Model}
We model the primary user's activity as a two-state discrete-time Markov chain. At any discrete time step $n$, let $X_n$ denote the state of the primary transmitter, where $X_n = 0$ represents an idle state (no transmission) and $X_n = 1$ represents an active state (primary is transmitting on the channel). The Markov chain is characterized by the following transition probabilities:
\begin{itemize}
    \item $P\{X_{n}=1 \mid X_{n-1}=0\} = \lambda$, the probability that the primary starts transmitting in the next time slot given that it was idle.
    \item $P\{X_{n}=0 \mid X_{n-1}=1\} = \mu$, the probability that the primary ceases transmission in the next time slot given that it was active.
\end{itemize}
From these, the complementary probabilities are $P\{X_{n}=0 \mid X_{n-1}=0\} = 1-\lambda$ (the primary remains idle) and $P\{X_{n}=1 \mid X_{n-1}=1\} = 1-\mu$ (the primary continues transmitting). This model assumes memoryless transitions dependent only on the current state, which is a reasonable approximation for many traffic patterns. The parameters $\lambda$ and $\mu$ can be estimated from historical observations of the primary's usage (e.g., based on average duty cycle and average burst length).

The steady-state behavior of this Markov chain yields the long-term fraction of time the primary is active or idle. In particular, the stationary probability that the primary is idle (state 0) is 
\[
\pi_0 = \frac{\mu}{\lambda + \mu},
\] 
while the stationary probability of the active state is $\pi_1 = \frac{\lambda}{\lambda + \mu} = 1 - \pi_0$. Intuitively, $\pi_0$ represents the expected availability of the channel (ignoring propagation effects) and is determined by the balance of the up and down transition rates.

\subsection{Propagation Model and Interference Check}
To incorporate spatial effects, we use propagation models to compute the path loss between the primary transmitter and a given secondary user receiver. We denote by $L_{\text{basic}}$ the basic transmission loss (in dB) predicted by ITU-R Recommendation P.528 \cite{ITU528}, and by $L_{\text{clutter}}$ the clutter loss (in dB) given by ITU-R Recommendation P.2108 \cite{ITU2108}. The basic loss $L_{\text{basic}}$ accounts for free-space loss, diffraction, tropospheric scatter, and other terrain-dependent effects for a path between the transmitter and receiver. P.528 is applicable for a wide range of frequencies (125~MHz to 15.5~GHz) and distances (up to several hundred kilometers) and requires inputs such as the distance between terminals, the heights of the antennas above sea level, the frequency, and a time availability factor (e.g., 95\% time). Using a time availability percentage (e.g., 95\%) allows the model to account for statistical variations in propagation (such as fading or ducting) by providing a loss value that is not exceeded for that percentage of time.

The clutter loss $L_{\text{clutter}}$ represents additional attenuation due to local obstructions around the receiver (such as buildings or trees in an urban or suburban environment). Recommendation P.2108 provides methods to calculate this loss, which can be added on top of the basic transmission loss. This model is typically applied when the receiver is at or below the average height of surrounding clutter and covers frequencies from 0.5~GHz to 67~GHz \cite{ITU2108}. The clutter loss can be estimated deterministically if detailed environment information is available (e.g., height of buildings, street widths), or statistically (providing a distribution of loss for a given environment type and percentage of locations).

By combining these components, the total path loss between the primary transmitter and secondary receiver is:
\begin{equation}\label{eq:loss}
L_{\text{total}} = L_{\text{basic}} + L_{\text{clutter}} \quad \text{(dB)}.
\end{equation}
The received power at the secondary user from the primary's transmission can be expressed (in dB) as:
\begin{equation}\label{eq:prx}
P_{\text{rx}} = P_{\text{tx}} + G_{t} + G_{r} - L_{\text{total}},
\end{equation}
where $P_{\text{tx}}$ is the primary transmitter's power (dBm or dBW), $G_{t}$ is the transmitter antenna gain (dBi), and $G_{r}$ is the receiver antenna gain (dBi). In this formulation, we assume that the secondary user's receiver is potentially directional or omnidirectional as characterized by $G_r$, and similarly for the primary transmitter.

Using \eqref{eq:prx}, we determine the channel availability condition for the secondary user. Let $P_{\text{th}}$ be the interference threshold (the maximum received power from the primary that the secondary can tolerate without significant interference). If $P_{\text{rx}} < P_{\text{th}}$, then the primary's signal is below the harmful interference level, and the channel can be considered available (provided the secondary transmits at a power that does not significantly raise the interference at the primary). On the other hand, if $P_{\text{rx}} \ge P_{\text{th}}$, the primary's signal is strong enough to interfere, and the channel is effectively occupied from the perspective of the secondary user.

It is worth noting that $L_{\text{basic}}$ and $L_{\text{clutter}}$ may be precomputed or approximated if the geometry and environment are known. For example, if the primary transmitter has a fixed location (or a known trajectory) and the secondary user's location is fixed, one could calculate $L_{\text{total}}$ once for that geometry (perhaps using a high percentile time availability to ensure a worst-case strong propagation scenario) and use that as a static parameter. In scenarios where either the primary or secondary is mobile, $L_{\text{total}}$ may change over time, and thus the interference check would be performed dynamically at each time step with updated distances or angles.

\subsection{Channel State Determination}
Combining the above elements, the instantaneous channel state (idle/available or busy/unavailable for secondary use) at a given secondary user location and time $n$ can be defined by the binary variable $Y_n$:
\[
Y_n = 
\begin{cases}
0, & \text{if channel is available at time $n$ (no interference)}; \\
1, & \text{if channel is occupied at time $n$ (primary interference present)}.
\end{cases}
\]
We can express $Y_n$ in terms of the primary activity $X_n$ and the interference check as:
\[ 
Y_n = 
\begin{cases}
1, & \text{if } X_n = 1 \text{ and } P_{\text{rx}}(n) \ge P_{\text{th}}, \\
0, & \text{otherwise}.
\end{cases}
\]
In other words, the channel is occupied ($Y_n=1$) for the secondary user only when the primary is actively transmitting \emph{and} that transmission is strong enough to be received above the threshold at the secondary's location. In all other cases (primary off, or primary on but signal too weak), $Y_n=0$ and the channel is effectively available.

Under the assumption that the propagation environment is static (or changes slowly relative to $X_n$), the condition $P_{\text{rx}} \ge P_{\text{th}}$ will consistently be true or false for a given location. For instance, if a secondary user is very close to the primary transmitter, any time the primary transmits we will have $P_{\text{rx}}$ exceeding $P_{\text{th}}$. Conversely, for a secondary user far away or heavily shielded by obstacles, it might be the case that $P_{\text{rx}}$ never exceeds $P_{\text{th}}$, effectively meaning this secondary never experiences interference from the primary (and $Y_n=0$ for all $n$). In intermediate situations near the edge of the primary coverage, occasional propagation enhancements (e.g., due to atmospheric effects captured by the time variability in P.528) might cause $P_{\text{rx}}$ to exceed $P_{\text{th}}$ at some times but not others. In such cases, one could extend the state model to include probabilistic occupancy when $X_n=1$ (for example, treating $Y_n$ as a probabilistic function of $X_n$ with a certain probability of interference if $X_n=1$). For simplicity, in our current framework we assume a deterministic threshold model: given the location and propagation model, we classify the secondary user as either within interference range or not for the primary's signal. Thus, $Y_n = X_n$ for locations within the interference range of the primary, and $Y_n = 0$ (always free) for locations outside the range.

The separation of concerns allows us to use the Markov chain $X_n$ to handle temporal prediction and the propagation model to handle spatial feasibility of interference. In the following section, we describe an algorithm that utilizes these models to predict $Y_n$ over time for one or multiple secondary user locations.

\section{Algorithm}
Using the above system model, we can formulate a step-by-step algorithm to predict spectrum availability for a given secondary user (or multiple users). The prediction can be performed either by Monte Carlo simulation of the Markov chain or by analytical computation of state probabilities. Here, we present a simulation-oriented pseudocode that iterates over time steps and updates the channel state.

\begin{algorithm}
\caption{Spectrum Availability Prediction Algorithm\label{alg:prediction}}
\begin{algorithmic}[1]
\State \textbf{Input:} Markov chain parameters $\lambda, \mu$; primary transmit power $P_{\text{tx}}$, antenna gains $G_t, G_r$; propagation model data (e.g. precomputed $L_{\text{total}}$ for location); interference threshold $P_{\text{th}}$; number of time steps $N$.
\State \textbf{Output:} Predicted availability timeline $\{Y_1, Y_2, \ldots, Y_N\}$ for the secondary user.
\State // Initialization
\State Generate initial primary state $X_0$ (idle or active) according to initial probabilities (e.g. stationary distribution $\pi_0, \pi_1$).
\For{$n = 1$ to $N$}
    \State // Evolve primary state using Markov chain
    \If{$X_{n-1} = 0$}
        \State $X_n \gets$ 1 with probability $\lambda$, or 0 with probability $(1-\lambda)$.
    \ElsIf{$X_{n-1} = 1$}
        \State $X_n \gets$ 0 with probability $\mu$, or 1 with probability $(1-\mu)$.
    \EndIf
    \State // Determine if channel is available at time $n$
    \If{$X_n = 1$} 
        \State Compute $L_{\text{total}}$ using propagation models (P.528 \& P.2108) if not precomputed.
        \State Compute $P_{\text{rx}} \gets P_{\text{tx}} + G_t + G_r - L_{\text{total}}$.
        \If{$P_{\text{rx}} < P_{\text{th}}$}
            \State $Y_n \gets 0$  \Comment{Channel is free (primary signal below threshold)}
        \Else 
            \State $Y_n \gets 1$  \Comment{Channel is occupied by primary}
        \EndIf
    \Else
        \State $Y_n \gets 0$  \Comment{Primary is idle, channel is free}
    \EndIf
\EndFor
\State \textbf{return} $\{Y_1, Y_2, \ldots, Y_N\}$
\end{algorithmic}
\end{algorithm}

In the pseudocode above, we simulate the primary user state $X_n$ over $N$ future time steps and, at each step, check if the channel is available (indicated by $Y_n = 0$) or occupied ($Y_n = 1$) for the secondary user. The propagation loss $L_{\text{total}}$ can be computed inside the loop if it varies with time (e.g., due to mobility or changing environmental conditions). If the geometry is static and $L_{\text{total}}$ is known a priori, it can be computed once before the loop and reused, which reduces computational load.

The algorithm as written predicts a specific sequence of channel states by effectively drawing samples from the stochastic process. In practice, one might run this simulation multiple times (or analytically compute probabilities) to estimate the likelihood of availability at each time. For instance, an analytical approach would update the probability $\Pr\{X_n = 1\}$ iteratively using the Markov chain transition matrix instead of drawing random states. That would produce a probability of the channel being busy or free at each time step, rather than a single sample path.

Nonetheless, the core steps remain the same: update the primary activity state according to the Markov chain, then apply the propagation and threshold check to determine channel status. This approach can be extended to multiple secondary users at different locations by repeating the interference check for each location (with potentially different $L_{\text{total}}$ values) at each time step. All users would share the same $X_n$ primary state in this scenario.

\section{Scalability and Efficiency}
A primary advantage of the proposed framework is its scalability to large numbers of secondary users and extended time horizons. We analyze the computational complexity and discuss how the algorithm can be efficiently implemented for real-world use.

In the simplest case, the time complexity of predicting spectrum availability for one secondary user over $N$ time steps is $O(N)$. Each time step involves a constant amount of work: updating the Markov chain state (which is $O(1)$) and performing a fixed number of arithmetic operations for the propagation loss and power calculations (also $O(1)$). If we extend the prediction to $M$ distinct secondary user locations (all relative to the same primary transmitter), the complexity becomes $O(N \times M)$, since for each time step we may repeat the interference calculation for each user. This linear scaling is quite manageable even for large $M$ or $N$. For example, if $N$ is on the order of $10^4$ time steps (perhaps representing real-time predictions for 10,000 future seconds) and $M$ is on the order of $10^3$ users, the total required operations are on the order of $10^7$, which is easily handled by modern processors within a fraction of a second to a few seconds depending on the operation cost.

Several factors further improve the efficiency:
\begin{itemize}
    \item \textbf{Propagation precomputation:} If the primary transmitter and secondary users are at fixed locations (or have a small set of discrete potential positions), one can precompute the path loss $L_{\text{total}}$ for each relevant geometry using P.528 and P.2108 before running the time iteration. The algorithm can then use stored $L_{\text{total}}$ values in the interference check, avoiding repeated calls to the propagation model. This is especially beneficial if the propagation model involves iterative or table lookup computations. In static scenarios, computing $L_{\text{total}}$ once per user (which might involve, say, evaluating a few equations from P.528 and P.2108) is a negligible overhead.
    \item \textbf{Parallelization:} The calculations for multiple secondary users at a given time step are independent of each other (assuming a single primary's state is broadcast to all, which it is). Therefore, those $M$ calculations can be parallelized across multiple CPU cores or threads. Similarly, if predictions for multiple channels (different primary transmitters or frequency bands) are needed, each channel's prediction can run in parallel. This embarrassingly parallel nature aligns well with modern multi-core processors and distributed computing, allowing near-linear speedup with additional computational resources.
    \item \textbf{Vectorized probability updates:} If an analytical approach is used instead of simulation, one can represent the state probability vector of the Markov chain and update it via matrix multiplication at each time step. This is a very light $2\times2$ matrix-vector multiplication in the two-state case, which is trivial to compute. Even if extended state models are considered, the state update can be vectorized and computed efficiently using linear algebra operations.
    \item \textbf{Memory usage:} The algorithm's memory footprint is minimal. It needs storage for the current state (or state probabilities) and the output array of length $N$ for $Y_n$ predictions (which could be streamed or processed on the fly if memory is constrained). For multiple users, storing one path loss value per user (or a small set of parameters per user) is sufficient. Thus, memory does not become a bottleneck even for large-scale predictions.
\end{itemize}

One potential concern for efficiency could be the execution of the propagation model itself. While P.528 and P.2108 are algebraic models, they have several conditional steps and calculations (for example, P.528 calculates horizon distances, checks line-of-sight conditions, and possibly transitions between different propagation modes like line-of-sight, diffraction, scatter, etc.). If these models are invoked repeatedly for many users or time steps, the constant factor in the complexity could grow. However, as noted, we can often mitigate this by precomputing or simplifying the model usage. Additionally, organizations like NTIA/ITS provide reference implementations of these models that are optimized in low-level languages; such implementations can compute path loss extremely quickly, and their results can be embedded into higher-level prediction code.

In summary, the Markov chain + propagation framework is highly scalable. Its runtime increases linearly with the number of predictions and locations, with low constant overhead. By leveraging precomputation and parallel processing, the method can support real-time operation in networks with numerous secondary users. This makes the approach well-suited not only for offline analysis of spectrum usage but also for live systems that continuously monitor and predict spectrum availability.

\section{Assumptions and Limitations}
While the proposed prediction framework is powerful, it relies on several assumptions and is subject to certain limitations that are important to recognize:

\begin{itemize}
    \item \textbf{Markovian traffic model:} We assume primary user behavior can be modeled as a memoryless two-state Markov chain with fixed transition probabilities $\lambda$ and $\mu$. In reality, primary traffic could have time-of-day patterns, long-term correlations, or multi-level activity states (e.g., varying transmission power or multiple transmitters on the same channel). If the primary exhibits non-Markovian behavior (such as periodic transmissions or bursty traffic with heavy-tailed off periods), the simple Markov chain might not capture the dynamics well. In such cases, a higher-order model, a semi-Markov process, or a time-varying $\lambda$, $\mu$ could be needed. The framework can be extended to incorporate these, at the cost of increased complexity.
    \item \textbf{Single primary transmitter:} Our model considers one primary transmitter as the source of potential interference. If a channel has multiple primary transmitters (for instance, co-channel transmitters in different locations, like several TV towers on the same frequency), the interference condition becomes a composite of signals from all of them. The Markov chain model could be expanded to include multiple primary states or multiple on/off processes, but the state space grows exponentially with the number of independent transmitters. A simplifying approach in those scenarios is to model an equivalent aggregated occupancy (the channel is busy if \emph{any} primary is transmitting and heard) which again reduces to a two-state model but with $\lambda$, $\mu$ reflecting the combined behavior. However, accuracy might suffer if primary transmitters are not homogeneous or independent.
    \item \textbf{Propagation model validity:} The use of ITU-R P.528 and P.2108 is appropriate for certain frequency ranges and scenarios (aeronautical/terrestrial path and clutter loss in urban/suburban areas, respectively). Applying these models outside their intended scope (e.g., P.528 for very low frequencies or purely ground-level links, or P.2108 in rural open areas without clutter) may reduce accuracy. Additionally, these models typically provide median or statistical loss estimates; actual instantaneous propagation can deviate due to fast fading or unusual atmospheric conditions. We partially account for this by using a conservative time percentage (like 95\%) in P.528, but this does not capture fast fading. The implication is that our prediction might occasionally declare a channel available while a deep fade temporarily made the primary undetectable, or vice versa.
    \item \textbf{Threshold and receiver sensitivity:} We assume a fixed interference threshold $P_{\text{th}}$ that cleanly divides interference vs no-interference. In practice, the impact of a primary signal on a secondary user can be gradual; a signal slightly below the threshold might still cause some performance degradation. Also, if the secondary user changes its own transmit power or sensitivity (for example, adapting transmission power), the effective threshold for harmful interference could change. Our model does not currently adapt to such changes, but it could be integrated with power control strategies by adjusting $P_{\text{th}}$ dynamically.
    \item \textbf{Lack of feedback/updating:} We present a prediction approach assuming $\lambda$, $\mu$, and propagation parameters are known and fixed. In a deployed system, it would be wise to continually update these parameters based on observations (e.g., if the primary's pattern changes or if a secondary user detects interference unexpectedly, that information could be used to refine the model). The framework allows for re-estimation of Markov transition probabilities over time, but we do not detail an adaptive mechanism here. Failing to update parameters in a non-stationary environment could lead to degraded prediction performance.
    \item \textbf{Secondary user impact:} We assume that the secondary users' activity does not influence the primary user's behavior (which is generally true if the primary is a licensed user oblivious to secondaries, as in typical spectrum sharing). We also assume secondary transmissions, when they occur, are configured not to cause significant interference to primary receivers (e.g., through obeying power limits and using the predicted availabilities). If these assumptions break (for example, secondary transmissions inadvertently affect the primary or if the primary has some sensing of secondaries), then a more complex coupled model would be required.
\end{itemize}

Despite these assumptions, the proposed model provides a solid baseline for many practical scenarios. Where necessary, it can be refined or supplemented with additional data. For instance, if real measurements of spectrum occupancy are available, they can be used to validate and adjust the Markov model. If more detailed propagation data (e.g., from ray tracing in a city) is available, it could replace or augment the ITU models for higher fidelity. The modular nature of our framework means that improvements in the temporal model or the spatial model can be incorporated independently.

\section{Applications}
The ability to predict spectrum availability in both time and space has numerous potential applications across wireless communications and spectrum management:

\begin{itemize}
    \item \textbf{Cognitive radio networks and dynamic spectrum access:} Secondary users in cognitive radio systems can use the prediction framework to schedule transmissions on channels that are likely to remain free, thereby reducing collision with primary users and improving throughput. For example, an IEEE 802.22 WRAN (Wireless Regional Area Network) device could predict TV channel availability before actually switching channels, to ensure stable operation.
    \item \textbf{Spectrum databases and policy enforcement:} Regulatory bodies or database-driven spectrum sharing systems (such as those used for TV white space or the 3.5~GHz CBRS band) can integrate predictive models to enhance their decision making. Rather than providing secondary users with static lists of available channels, a database could offer predictive availability (with confidence metrics) to help devices choose frequencies that will likely remain available for the duration of their transmission. This can improve quality of service for secondary users while still protecting primary users.
    \item \textbf{Wireless network planning and deployment:} When deploying secondary networks (e.g., sensor networks or IoT devices that opportunistically use spectrum), planners can use this model to assess the expected availability of channels in different locations and times. This informs decisions like how often devices can transmit, what data rates can be sustained, or what backup links to have if a channel becomes unavailable.
    \item \textbf{Interference management for aerial and satellite systems:} In scenarios where primary users are airborne or satellite transmitters, the Markov chain with P.528 propagation is directly applicable. For instance, a ground network sharing spectrum with aeronautical communications can predict when an aircraft (serving as a primary transmitter) will cause interference at a given ground location as it flies (since P.528 handles air-to-ground paths). The ground network can then proactively avoid using the frequency during those periods. Similarly, for non-geostationary satellites that appear and disappear from view, a predictive model can schedule transmissions in the gaps when the satellite is not overhead or not transmitting.
    \item \textbf{Adaptive communication protocols:} Protocols that adapt their behavior based on spectrum conditions (such as frequency hopping spread spectrum or multi-band LTE/5G systems) can incorporate our prediction engine to decide which band to hop to or which carrier to activate next. By predicting future interference, the protocol can make more informed choices, leading to fewer retransmissions and lower latency.
\end{itemize}

These examples illustrate the broad utility of spectrum availability prediction. As wireless ecosystems become more complex with spectrum sharing between different services (e.g., radar and communications, or incumbents and new entrants), having a robust predictive tool will be essential to manage coexistence. The framework we propose is general enough to be tailored to many such scenarios by selecting appropriate propagation models and adjusting the Markov chain parameters to the traffic characteristics of the primary system in question.

\section{Conclusion}
In this paper, we presented a comprehensive framework for predicting spectrum availability that integrates a Markov chain model of primary user activity with ITU-R propagation models to account for spatial signal attenuation. This combination allows us to predict not only when a primary user is likely to transmit, but also where those transmissions will actually pose interference to secondary users. Our proposed model maintains technical accuracy by leveraging established propagation prediction methods (P.528 for basic path loss and P.2108 for clutter loss) and balances that with the simplicity and scalability of a two-state Markov chain.

We provided a detailed system model, pseudocode algorithm, and analysis of the method's computational efficiency. The framework is shown to be scalable to large networks and fine-grained time predictions, largely because of its linear complexity and ability to use precomputed propagation data. We also discussed the assumptions under which the model operates and its limitations, ensuring that its applicability to real-world scenarios is well-understood.

The proposed approach can serve as a building block for more intelligent dynamic spectrum access systems, cognitive radios, and spectrum management tools. By proactively predicting when and where spectrum will be free, secondary users can make better decisions and coexist more harmoniously with primary users. 

\paragraph{Future Work:} 
There are several avenues for extending this work. One direction is to incorporate machine learning techniques to adjust the Markov model parameters on the fly or to capture non-Markovian patterns in primary traffic. Another extension is to consider multiple primary transmitters or multiple channels simultaneously, possibly using multi-dimensional state models or factorized approximations to keep complexity tractable. Additionally, extensive field testing and validation of the model with real spectrum measurement data would be valuable to quantify prediction accuracy and refine the propagation components under various conditions. Finally, integration with actual communication systems (e.g., implementing this prediction in a cognitive radio's spectrum access protocol) would demonstrate its practical benefits in reducing interference and improving spectral efficiency.

\bibliographystyle{unsrt}

\end{document}